\documentclass[twocolumn,aps,prl,amsmath,amssymb,superscriptaddress,bibnotes]{revtex4-1}
\usepackage{amssymb}
\usepackage{natbib}
\usepackage{amsmath}
\usepackage{amsfonts}
\usepackage{graphicx}
\usepackage{mathrsfs}
\usepackage{dcolumn} 
\usepackage{bm}
\usepackage{color}
\usepackage[colorlinks=true, citecolor=blue, urlcolor=blue]{hyperref}
\usepackage{cancel}
\usepackage{mathbbol}
\usepackage{epsfig}
\usepackage{units}
\usepackage{esint}
\usepackage{soul} 
\usepackage{hyperref,soul}
\usepackage{float}
\usepackage{diagbox}
\usepackage{setspace}

\begin{document}

\title{Experimental demonstration of quantum cooling engine powered by entangled measurement}

\author{Ning-Ning Wang}
\affiliation{Key Laboratory of Quantum Information, University of Science and Technology of China, CAS, Hefei, 230026, People's Republic of China}
\affiliation{CAS Center For Excellence in Quantum Information and Quantum Physics, University of Science and Technology of China, Hefei, 230026, People's Republic of China}
\affiliation{Hefei National Laboratory, University of Science and Technology of China, Hefei 230088, China}

\author{Huan Cao}
\affiliation{Key Laboratory of Quantum Information, University of Science and Technology of China, CAS, Hefei, 230026, People's Republic of China}
\affiliation{University of Vienna, Faculty of Physics, Vienna Center for Quantum Science and Technology (VCQ) and Research platform TURIS, Boltzmanngasse 5, 1090 Vienna, Austria}

\author{Chao Zhang}
\email{zhc1989@ustc.edu.cn}
\affiliation{Key Laboratory of Quantum Information, University of Science and Technology of China, CAS, Hefei, 230026, People's Republic of China}
\affiliation{CAS Center For Excellence in Quantum Information and Quantum Physics, University of Science and Technology of China, Hefei, 230026, People's Republic of China}
\affiliation{Hefei National Laboratory, University of Science and Technology of China, Hefei 230088, China}

\author{Xiao-Ye Xu}
\affiliation{Key Laboratory of Quantum Information, University of Science and Technology of China, CAS, Hefei, 230026, People's Republic of China}
\affiliation{CAS Center For Excellence in Quantum Information and Quantum Physics, University of Science and Technology of China, Hefei, 230026, People's Republic of China}
\affiliation{Hefei National Laboratory, University of Science and Technology of China, Hefei 230088, China}

\author{Bi-Heng Liu}
\affiliation{Key Laboratory of Quantum Information, University of Science and Technology of China, CAS, Hefei, 230026, People's Republic of China}
\affiliation{CAS Center For Excellence in Quantum Information and Quantum Physics, University of Science and Technology of China, Hefei, 230026, People's Republic of China}
\affiliation{Hefei National Laboratory, University of Science and Technology of China, Hefei 230088, China}

\author{Yun-Feng Huang}
\email{hyf@ustc.edu.cn}
\affiliation{Key Laboratory of Quantum Information, University of Science and Technology of China, CAS, Hefei, 230026, People's Republic of China}
\affiliation{CAS Center For Excellence in Quantum Information and Quantum Physics, University of Science and Technology of China, Hefei, 230026, People's Republic of China}
\affiliation{Hefei National Laboratory, University of Science and Technology of China, Hefei 230088, China}

\author{Chuan-Feng Li}
\email{cfli@ustc.edu.cn}
\affiliation{Key Laboratory of Quantum Information, University of Science and Technology of China, CAS, Hefei, 230026, People's Republic of China}
\affiliation{CAS Center For Excellence in Quantum Information and Quantum Physics, University of Science and Technology of China, Hefei, 230026, People's Republic of China}
\affiliation{Hefei National Laboratory, University of Science and Technology of China, Hefei 230088, China}

\author{Guang-Can Guo}
\affiliation{Key Laboratory of Quantum Information, University of Science and Technology of China, CAS, Hefei, 230026, People's Republic of China}
\affiliation{CAS Center For Excellence in Quantum Information and Quantum Physics, University of Science and Technology of China, Hefei, 230026, People's Republic of China}
\affiliation{Hefei National Laboratory, University of Science and Technology of China, Hefei 230088, China}

\begin{abstract}
    Traditional refrigeration is driven either by external force or an information-feedback mechanism. Surprisingly, the quantum measurement and collapse, which are generally detrimental, 
    can also be used to power a cooling engine even without requiring any feedback mechanism. In this work, we experimentally demonstrate quantum measurement cooling (QMC) powered by entangled measurement 
    by using a novel linear optical simulator. In the simulator, different thermodynamic processes can be simulated by adjusting the energy-level spacing of working substance and the temperature of thermal bath. We show experimentally that, without prior knowledge 
    about the measurement to be made, QMC remains likely to occur. We also demonstrate that QMC is robust against measurement noise. Those experimental results show that quantum measurement 
    is not always detrimental but can be a valuable thermodynamic resource.
\end{abstract}

\maketitle

    {\it Introduction.--} Quantum measurement is one of the most peculiar concepts of quantum mechanics. Contrary to its classical counterpart, the wave function of the measured system will collapse to an eigenstate of the observable with a certain probability. 
    Von Neumann describes the quantum measurement process to be an interaction between system and measurement apparatus \cite{von}, the combined system evolves unitarily according to the Schr\"{o}dinger equation while the evolution of the system alone is nonunitary. 
    Unlike some other quantum effects such as coherence \cite{coherence1,coherence2,coherence3,coherence4,coherence5}, quantum correlation \cite{entanglement1,entanglement2,cor1,cor2}, or squeezed bath \cite{squeezed1,squeezed2,squeezed3}, acting as enhancement roles in some thermodynamic tasks, 
    the quantum measurement is generally thought to be detrimental \cite{backaction1,backaction2} because its induced backaction destroys the quantum coherence of measured system 
    between different measurement bases. However, this genuine quantum effect can also be helpful in some cases, e.g. to fuel a quantum engine, which indicates a quantum measurement can also be viewed as a useful thermodynamic resource \cite{res}. 

    Previously, discussions on the role of measurement in thermodynamic engines have focused mostly on the aspects of extracting information \cite{inf1,inf2,inf3,inf4}. For example, in the mechanism of Maxwell's demon, information gained from the measurement of the working substance can be used to design appropriate feedback to achieve thermodynamic tasks
    \cite{demon1,demon2,demon3,demon4,demon5,demon6,demon7,demon8}. In this scenario, the thermal bath is usually required to play the role of energy source. However, in the quantum scenario, the measurement can be invasive, meaning that energy exchange between the working substance and 
    measurement apparatus occurs when the observable is not commutative with the Hamiltonian of working substance. Specifically, the post-measurement state can contain more energy than its initial state. In this sense, the measurement itself can also be 
    an energy source. Based on this principle, heat engines with feedback (quantum Maxwell's demon) or without feedback (ignore the measurement results) have been proposed \cite{QEM1,QEM2,QEM3,QEM4,QEM5,QEM6,QEM7,QEM8}. In these engines, working substance no longer interacts with hot baths during the energizing strokes but gains 
    energy from invasive quantum measurement directly. 

    A natural question is whether all measurement processes can increase the average energy of a system. The answer is definite for a passive state which contains no ergotropy \cite{QEM6}. For example, a single qubit working substance in equilibrium with a thermal bath always gains energy from the measurement process. 
    However, things become different when considering how an entangled measurement affects subsystems due to its inseparability. Even if the subsystem is in a passive state, its energy can be lowered during the entangled measurement. 

    Lately, a quantum mechanical refrigeration concept--quantum measurement cooling (QMC)--was proposed \cite{QMC1,QMC2}. QMC is implemented in a two-stroke two-qubit cooling engine fueled only by entangled measurement. 
    The collective measurement reduces the energy of the qubit in thermal equilibrium with a cold bath, thus allowing this qubit to take heat away from the cold bath during the engine cycle. But overall, the quantum measurement provides energy for the reverse flow of heat. 
    Interestingly, this protocol requires no information on measurement results and no feedback operations; the demon that performs the quantum measurement only needs to know which measurement basis can be used for refrigeration. 
    This means that QMC is robust against experimental noise. Despite its robustness, however, the significant energy level spacing between two qubits required in QMC protocol is challenging, so such a proposal has yet to be realized \cite{QMC2}. 

    In this Letter, we verify QMC experimentally in a two-qubit two-stroke quantum measurement cooling engine constructed by a linear optical simulator. 
    First, by using various working substances with different energy levels, we verify the predicted QMC and three other thermodynamic operations that can be implemented by the cooling engine. Next, in the spirit of 
    no prior knowledge (i.e., random measurement settings), we evaluate the QMC frequency. Finally, we show that this refrigeration process is robust against noise. 
        
    \begin{figure}
        \includegraphics[scale = 0.5]{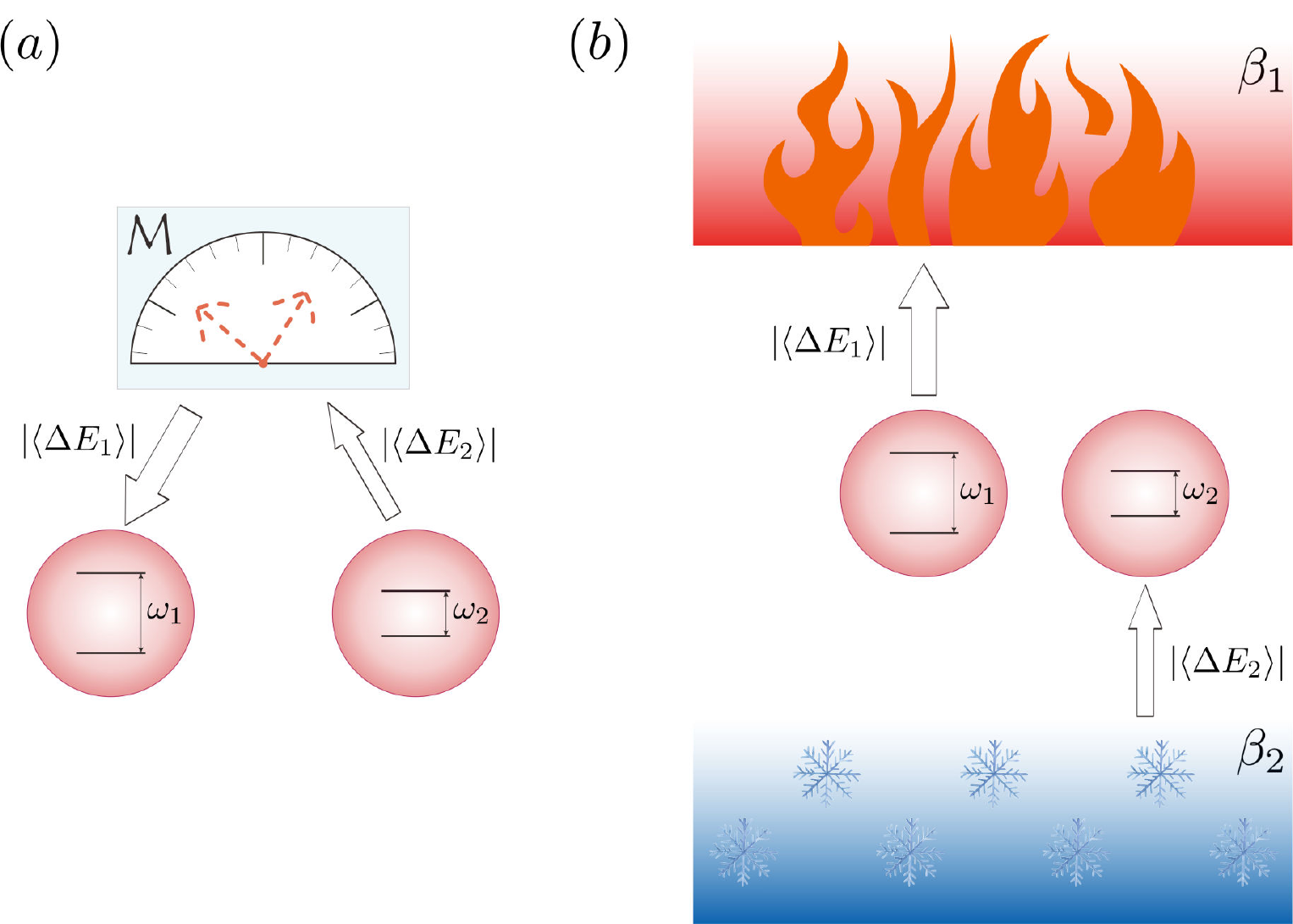}
        \caption{Two strokes of the cooling engine. The arrows show the direction of energy transfer. (a) The first stroke. The working substance interacts and exchanges energy with 
            the measurement apparatus during the measurement process, but the measurement results are not read. (b) The second stroke. The collapsed 
            working substance contacts the thermal baths and reaches equilibrium to close the engine cycle.}
        \label{cycle}
    \end{figure}

    {\it Theory and model.--} Considering a two-qubit working substance with total Hamiltonian $\mathit{H}  = \mathit{H}_1 + \mathit{H}_2$, where $\mathit{H}_i = -\omega_i\sigma_z^i/2$ 
    is the Hamiltonian of qubit $i$ expressed by the Pauli matrix $\sigma_z^i$ and $\omega_i > 0$ is the energy-level spacing (in units of $\hbar\cdot\mathrm{Hz}$), its energy eigenstates can be written as $|0_i\rangle$ and $|1_i\rangle$, 
    corresponding to the energy eigenvalues $-\omega_i/2$ and $\omega_i/2$, respectively. 
    The system has two thermal baths with inverse temperatures $\beta_1$ and $\beta_2$ (in units of $k_b^{-1}\cdot\mathrm{K}^{-1}$), where $0 < \beta_1 < \beta_2$, which means that thermal bath $1$ is hotter than thermal bath $2$.
    Initially, each of the two qubits makes contact with its corresponding thermal bath to reach an equilibrium state; their joint state is denoted $\rho = \rho_{1}\otimes \rho_{2}$, where
    \begin{equation}
        \rho_{i} = \frac{1}{\mathit{Z}_i} e^{-\beta_i\mathit{H}_i}, \quad  \mathit{Z}_i = {\rm Tr}(e^{-\beta_i\mathit{H}_i})
        \label{equi}
    \end{equation}
    $\mathit{Z}_i$ is the partition function. The energy of each qubit is perfectly anti-correlated with its corresponding population, 
    but $\rho$, which is the direct product of $\rho_1$ and $\rho_2$, is probably not this case because the two qubits have different energy-level spacings and their thermal 
    equilibrium temperatures also differ. 

    The cooling engine cycle contains two strokes, see Fig.~\ref{cycle}. In the first stroke, the working substance interacts with the measurement apparatus and is projected into a set of two-qubit bases 
    $\{|\psi_k\rangle\}_{k = 1 }^{4}$. Compared with Maxwell's demon, no information about the measurement results is gained. That is, 
    we do not assume post-selection on the measurement output or any feedback control. Thus the post-measurement state $\rho^\prime$ becomes the mixture of measurement bases: 
    $\rho^\prime = \sum_{k = 1}^{4}\langle\psi_k|\rho|\psi_k\rangle|\psi_k\rangle\langle\psi_k|$. During the interaction between working substance and measurement apparatus, 
    energy exchanges, with the net energy flow being $\langle \Delta E\rangle = \langle \Delta E_1\rangle + \langle \Delta E_2\rangle$, where $\langle \Delta E_i\rangle = {\rm Tr} ((\rho^\prime - \rho) \mathit{H}_i)$ 
    refers to the energy change of qubit $i$. Meanwhile, the second law of thermodynamics compels $\beta_1\langle \Delta E_1 \rangle + \beta_2 \langle \Delta E_2 \rangle \ge 0$ \cite{2thlaw}. 

    In the second stroke, qubit 1 (2) makes contact with thermal bath 1 (2) to reach thermal equilibrium again. This exchanges energy between the working substance 
    and the thermal baths, which is equivalent to a cooling process transferring heat from cold bath 2 to hot bath 1. This completes the initialization for the next cycle.

    Constrained by conservation of energy and the second law of thermodynamics, four operations are possible \cite{QMC1}: 
    (1) $[R]$, refrigeration, in which $\langle \Delta E_1 \rangle \ge 0, \langle \Delta E_2 \rangle \le 0, \langle \Delta E \rangle \ge 0$, and the working substance gains energy from the measurement apparatus and thereby transports thermal energy from the cold bath to the hot bath. 
    (2) $[E]$, energy extraction, in which $\langle \Delta E_1 \rangle \le 0, \langle \Delta E_2 \rangle \ge 0, \langle \Delta E \rangle \le 0$, and energy flow from the hot bath to the cold bath while being partially extracted by the measurement apparatus. 
    (3) $[A]$, thermal acceleration, in which $\langle \Delta E_1 \rangle \le 0, \langle \Delta E_2 \rangle \ge 0, \langle \Delta E \rangle \ge 0$, and more thermal energy flow from the hot bath to the cold bath with the energy injection from the measurement apparatus, as compared with the natural flow of heat. 
    (4) $[H]$, heater, in which $\langle \Delta E_1 \rangle \ge 0, \langle \Delta E_2 \rangle \ge 0, \langle \Delta E \rangle \ge 0$, and the measurement apparatus heats both thermal baths by consuming its energy. 
    Note that QMC is related to $[R]$, which could occur only if some of the measurement projectors are entangled. 
    
    \begin{figure}
        \centering
        \includegraphics[scale = 0.089]{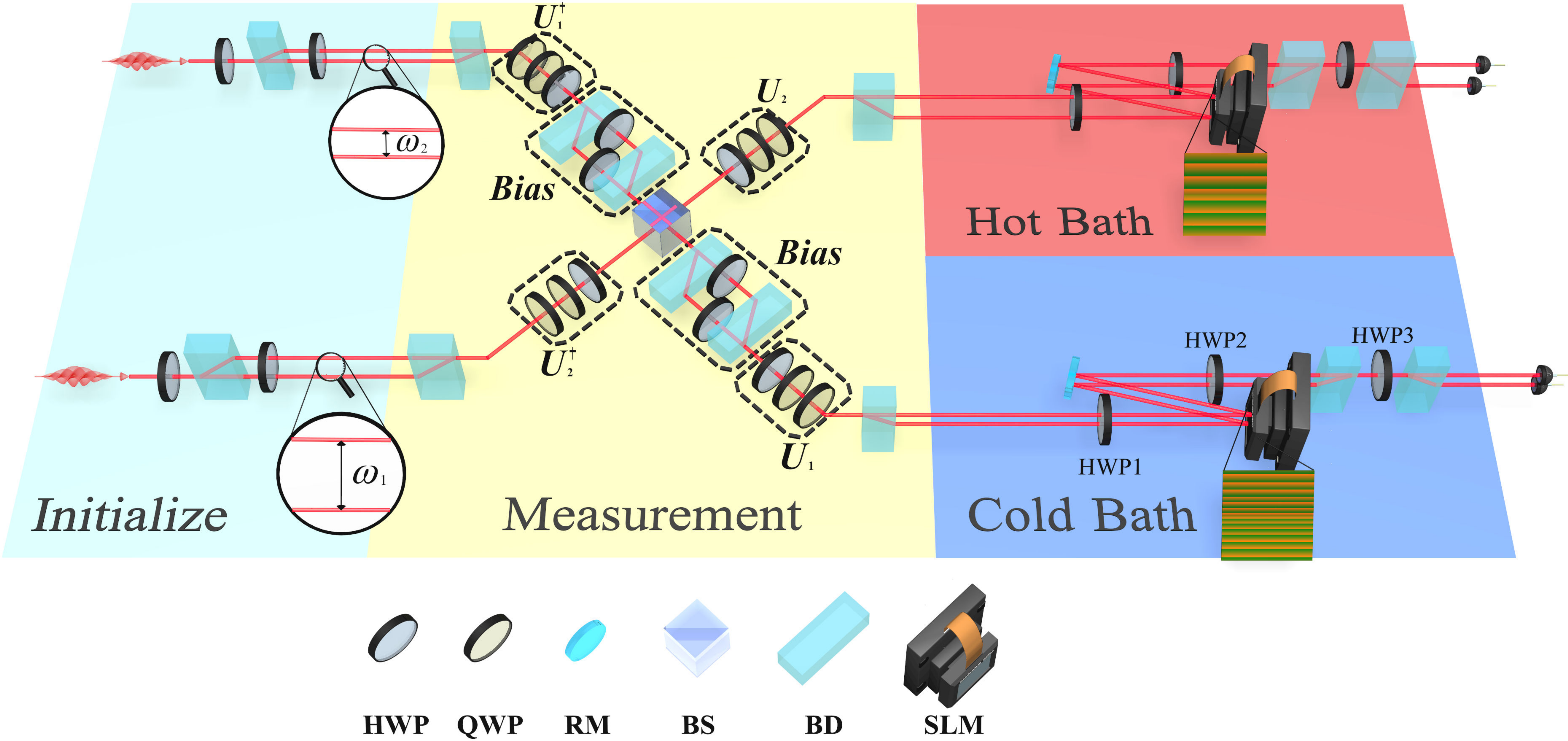}
        \caption{Experiment setup. The component abbreviations are HWP, half-wave plate; QWP, quarter-wave plate; RM, reflective mirror; 
        BS, beam splitter; BD, beam displacer; SLM, spatial light modulator. The simulator contains three parts: (a) Initialize (cyan region). 
        Depending on the energy-level spacings of the two qubits to be simulated, two different BDs are used in the setup. 
        The BD does not change the path of vertically polarized photons but translates horizontally polarized photons downward by a distance that depends on the thickness of BD. The HWP placed behind the BD is set to $45^o$ to complete the initialization. 
        (b) Measurement (yellow region). BDs are used to combine photons from different paths into one path or vice versa. Three cascaded wave plates are used to implement arbitrary polarization unitary transformation. The unitaries before and after the BS 
        are Hermitian with respect to each other along each path. A Mach-Zehnder (M-Z) interferometer is composed of two BDs and two HWPs on each arm. Two M-Z interferometers are arranged symmetrically with respect to the BS on one path. 
        (c) Thermalizing channel ($\beta_1$ in the red region and $\beta_2$ in the blue region). The hologram displayed on the SLM reflects the temperature of the thermal bath. 
        All the wave plates are mounted on motorized computer-controlled rotation stages. }
        \label{setup}
    \end{figure}

    {\it Linear-optical simulator.--} The optical version of QMC is now described (see Fig.~\ref{setup}). 

    (1) {\it Initialize.} The two-qubit working substance is simulated by twin photons generated by spontaneous parametric down-conversion.
    Specifically, the ground (excited) state $|0_i\rangle$ ($|1_i\rangle$) of qubit $i$ is encoded in photonic path-polarization mode 
    $|H\rangle_{\textnormal{-}d_i/2}$ ($|V\rangle_{d_i/2}$), where $|H\rangle_{\textnormal{-}d_i/2}$ ($|V\rangle_{d_i/2}$) denotes the state of photon $i$ 
    with horizontal (vertical) polarization at the position $\textnormal{-}d_i/2$ ($d_i/2$), and the energy level spacing of the qubit is simulated by the spatial spacing $d_i$ of the two states \cite{supp}. 
    Applying a beam displacer (BD) with suitable thickness, the qubit with arbitrary energy level spacing can, in principle, be prepared. 
    The initial state $\rho$ of the working substance is prepared by time-mixing its ground and excited states, and the time intervals are set proportionally according to the population of energy levels. 

    (2) {\it Measurement.} The measurement process is a vital ingredient of the cooling engine. As a demonstration in the photonic testbed, we convert the measurement of energy space into the photonic polarization space. 
    This is done by inserting BDs of the same specifications as in the initialization part. 

    In the spirit of removing access to any information of measurement output, we mix four projective operations of a measurement process classically. The two-qubit polarization projection can be achieved by the singlet projection and local operations \cite{projector,supp}. 
    In our setup, the singlet projector is realized with a beam splitter (BS), at which the twin photons meet and undergo Hong-Ou-Mandel (HOM) interference. By post-selecting the case of each output port containing one photon, 
    two photons are projected to a polarization singlet state. The local operations are local unitary transformations that are performed with QWP-QWP-HWP waveplate structures, and polarization-dependent adjustable coefficients modules 
    which are based on the Mach-Zehnder interferometer (the dotted boxes labeled {\it Bias} in Fig.~\ref{setup}). 
    Finally, the two qubits are restored to their energy space by using the same BDs as before. We thus complete the first stroke of our cooling engine.

    To verify the validity of our setup, we performed the quantum measurement tomography \cite{meato} on several randomly selected bases, with an average fidelity of about 0.98, verifying the implementation of the proposed measurement processes \cite{supp}. 

    (3) {\it Thermalizing.} The interaction between a two-level qubit and a thermal bath can be modeled by a generalized amplitude 
    damping channel (GAD) which contains four Kraus operators$\{K_1, K_2, K_3, K_4\}$ \cite{channel1,channel2,channel3}. 
    As the interaction time increases and tends to be infinite qubit reaches thermal equilibrium with its thermal bath, and GAD is referred to as the thermalizing channel. 

    The red and blue region in Fig.~\ref{setup} shows the implementation of the thermalizing channel. As an example, let's consider the qubit contacting with a cold bath. 
    The core device is a spatial light modulator (SLM). The imprinted phase by SLM is related to the position $z$, and its distribution, i.e. $\phi_{\beta_2}(z)$, reflects the 
    temperature $\beta_2$ of the cold bath. When the photonic qubit with two path modes enters the cold bath and is incident on the $\pm d/2$ position of SLM, these two 
    modes will evolve differently according to their path position and translate into each other \cite{supp}. This process is like the transition between two energy levels of the working substance when contacting with a thermal bath, 
    thus simulating the interaction between working substance and cold bath. Operationally, two settings with specific angles of (HWP1, HWP2, HWP3) corresponding 
    to two pairs $\{K_1, K_3\}$ or $\{K_2, K_4\}$ of Kraus operators are switched randomly to realize the Kraus representation of thermalizing channel.
    
    Process tomography of the thermalizing channel is also performed \cite{proto}, for which we choose the baths at different temperatures and working substances with different energy gaps. 
    The result shows an average fidelity exceeding $0.99$, verifying the high performance of our channel simulation \cite{supp}.

\begin{figure}
        \centering
        \includegraphics[scale = 1]{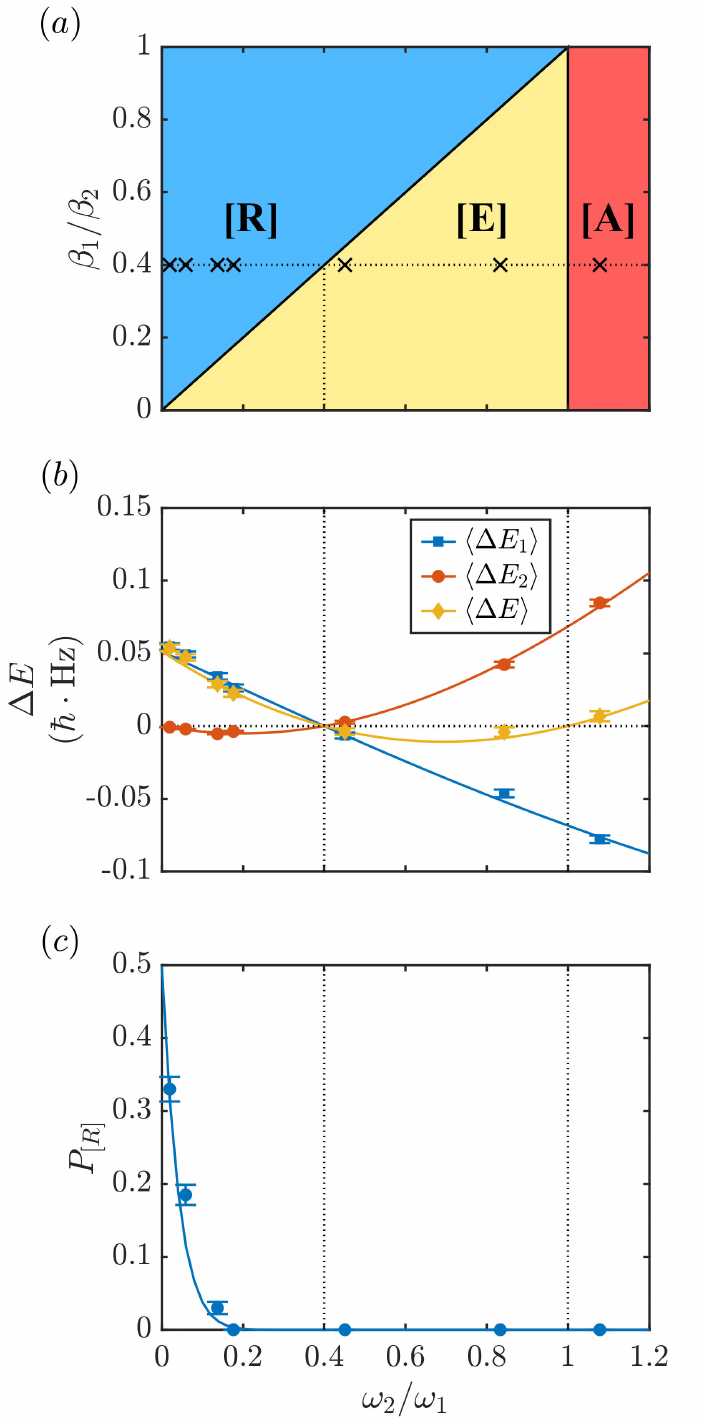}
        \caption{(a) Parameter regions. The parameters used in our experiment are marked with crosses. 
        (b) Different thermodynamic operations of our measurement fueled engine when using working substances with different energy spacings.  
        (c) The frequency of successful realization of the [R] operation, when the measurement is performed randomly. The solid line is a result of Monte Carlo sampling and the blue circles are experimental results.
        All the quoted errors are derived from Poissonian statistics on the raw data.}
        \label{exresult}
    \end{figure}

    {\it Experiment result.--} The first result is that we verify the operations allowed to realize during the engine cycle, in terms of working substance with different energy level spacings. 
    The parameter ranges of bath temperatures and working substance energy level spacings can be divided into three classes: 
    For $0 \le \omega_2/\omega_1 \le \beta_1/\beta_2$, [R]-range, only [R] and [H] are allowed. For $\beta_1/\beta_2 \le \omega_2/\omega_1 \le 1$, [E]-range, only [E], [A], 
    and [H] are allowed. For $\omega_2/\omega_1 \ge 1$, [A]-range, only [A] and [H] are allowed. Fig.~\ref{exresult}(a) plots the regions of these three different ranges with different colors. As a proof-of-principle demonstration, we prepare the baths with inverse temperature 
    fixed at $\beta_1 = 1/2.5$ and $\beta_2 = 1$ by embedding the corresponding hologram $\phi_{\beta_i}(z)$ on the SLM. Moreover, we fix the energy gap of qubit 1 at $\omega_1 = 1.02$ and vary the 
    energy-level spacings of the second qubit over the set of $\omega_2 = \{0.02, 0.06, 0.14, 0.18, 0.46, 0.86, 1.10\}$ by replacing BDs with different deflecting distances. This parameter configuration corresponds to 
    a horizontal line with $\beta_1/\beta_2 = 0.4$ in Fig.~\ref{exresult}(a). The measurement consists of four projector bases $\{|\psi_k\rangle\}$, with two of them maximally entangled: 

    \begin{equation}
        \begin{aligned}
            |\psi_1 \rangle &= |00\rangle, &|\psi_2 \rangle &= \frac{1}{\sqrt{2}}(|01\rangle + |10\rangle),\\
            |\psi_3 \rangle &= \frac{1}{\sqrt{2}}(|01\rangle - |10\rangle), &|\psi_4 \rangle &= |11\rangle 
        \end{aligned}
        \label{basis}
    \end{equation}

    This measurement can be verified to be invasive and not separable. After the first stroke, 
    the energy of $\rho^\prime$, which is the post-measurement state, is recorded, because the energy exchange occurs during the interaction between the working substance and the measurement apparatus. In the second stroke, 
    the working substance makes contact with the thermal bath to exchange energy and reach equilibrium, which terminates the engine cycle. 
    The energy of the equilibrium state is measured and compared with the energy of $\rho^\prime$ to determine which process occurs. 

    For each $\omega_2$ we calculate $\langle\Delta E\rangle$, $\langle\Delta E_1\rangle$, and $\langle\Delta E_2\rangle$, 
    and the results are shown in Fig.~\ref{exresult}(b). Three different processes are possible according to different $\omega_2$: as $\omega_2$ increases, 
    $[R]$, $[E]$, and $[A]$ are successively realized. In the $[R]$ range, $\left\rvert\langle \Delta E_2 \rangle\right\rvert$, which is the 
    heat extracted from the cold bath, does not monotonically increase as $\omega_2$ increases. However, $\left\rvert\langle \Delta E \rangle\right\rvert$, 
    which is the pump energy provided by the measurement apparatus, decreases as $\omega_2$ increases, which indicates a greater refrigeration efficiency when $\omega_2/\omega_1$ reaches $\beta_1/\beta_2$. 
    In the $[E]$ range, $\rho$ is not a passive state, that is, the energy-level population is not inversely related to its energy. Some measurement processes, therefore, exist 
    that can extract energy from working substances. We see that the energy $\left\rvert\langle \Delta E \rangle\right\rvert$ flow to the measurement apparatus, first increasing, and then decreasing to zero at $\omega_1 = \omega_2$. 
    However, the energy $\left\rvert\langle\Delta E_1\rangle\right\rvert$ extracted from the hot bath is always increasing, which means that a small energy-level spacing for qubit 2 makes the operation $[E]$ more efficient. 
    The last non-trivial process $[A]$ occurs when $\omega_2 \ge \omega_1$. The larger $\omega_2$ is, the faster the energy exchange occurs between the hot bath and the cold bath.

    The second result is that the frequency $P_{[R]}$ of the QMC is realized with no prior knowledge of what measurement to do, illustrating the robustness of QMC. Experimentally, 
    the measurement basis is randomly selected. This is done by randomly picking several unitaries $U$ from the SU(4) invariant (Haar) measure. The unitaries are then applied to the projectors in Eq.~\eqref{basis}, 
    producing the new basis $\{U|\psi_k\rangle\}_{k=1}^{4}$. We generated 200 unitary matrices from Haar measures with a computer and implement 200 measurement settings for each type of working substance. 
    During the engine cycles, we measure $\langle\Delta E\rangle$, $\langle\Delta E_1\rangle$, 
    and $\langle\Delta E_2\rangle$ and calculate the direction of heat flow to determine whether QMC occurs for each measurement setting. 
    Fig.~\ref{exresult}(c) shows that QMC is possible only when $\omega_2/\omega_1 \le \beta_1/\beta_2$. The frequency of the $[R]$ operation increases and tends to $0.5$ as 
    $\omega_2/\omega_1$ decreases and tends to zero, showing that QMC becomes more robust against the randomness of measurement with decreasing $\omega_2/\omega_1$. These results also indicate that 
    a significant difference in energy level spacings between the two qubits is necessary to implement QMC. If we average all the measurement effects, the most useless operation $[H]$ is implemented \cite{supp}.

    Finally, note that the realization of QMC is irrelevant to the measurement efficiency. During the measurement process, we consider the measurement efficiency of $\nu$, meaning that the measurement apparatus interacts 
    with the working substance with probability $\nu$, while $1-\nu$ is the probability of no interaction. No matter how inefficient the measurement process is, the QMC can still be realized because the energy of the working substance does 
    not change if it does not interact with the measuring apparatus--only a successful measurement contributes to the energy exchange. In addition, the experimental noise of our simulator is mainly due to imperfect HOM interference. 
    Although this may eventually prevent the refrigerator from working, we find that our setup remains noise-tolerant, especially when the energy-level spacing between two qubits is significant (see the Supplemental Material for more details \cite{supp}).

    {\it Conclusion.--} In summary, we propose herein a simulation protocol in which the energy level of working substance and the temperature of the thermal baths are easily adjusted. 
    In addition, we experimentally demonstrate QMC with a quantum cooling engine built by this simulator, which means that we transfer heat from the cold bath to the hot bath by pumping with an invasive quantum entangled measurement. 
    We show that this type of measurement-fueled cooling engine is robust against experimental noise, and the results are consistent with the theory. 
    This evidence indicates that quantum measurements can be a useful thermodynamic resource. 
    
    Although the thermodynamic quantities obtained are only from the simulation of an actual situation \cite{channel3,thermometer1,otto,simu}, 
    our experiment still deepens our understanding of quantum measurements and provides a convenient setup for future research into quantum thermodynamics, such as measurement-based quantum engines \cite{QEM1,QEM2,QEM3,QEM4,QEM5,QEM6,QEM7,QEM8}, 
    environment-system interaction \cite{thermometer2,thermometer3,chan1,chan2}, quantum algorithm \cite{alg1,alg2}, and so on.

    {\it Acknowledgments.--} We thank Sheng Liu for the beneficial discussions. This research was supported by the Innovation Program for Quantum Science and Technology (No. 2021ZD0301604), the National Natural Science Foundation of China (Nos. 11821404, 11734015, 62075208, 12022401, 62075207). 
    This work was partially carried out at the USTC Center for Micro and Nanoscale Research and Fabrication.

    \clearpage

    \begin{widetext}
        \section*{Supplemental Material}

        \renewcommand{\thefigure}{S\arabic{figure}}
        \renewcommand{\theequation}{S\arabic{equation}}
        \renewcommand{\thetable}{S\arabic{table}}
        \setcounter{equation}{0}
        \setcounter{figure}{0}
        \setcounter{table}{0}

        \subsection{1. Simulation of working substance} 
        In this section, we introduce the simulation protocol of the energy level of the working substance in detail. As described in the main text, we use the photonic path-polarization mode 
        $|H\rangle_{\textnormal{-}d/2}$ and $|V\rangle_{d/2}$ to encode the ground state $|0\rangle$ with the energy of $-\omega/2$ and the excited state $|1\rangle$ with the energy of $\omega/2$ respectively. 
        Practically, since both the photonic beam and the pixels on the SLM are of a certain size, the energy level spacing $\omega$ is set to have the same value when photonic qubits incident in a certain area, see Fig.~\ref{fig_S1}(a). 
        The specification of the SLM is 512 $\times$ 512 pixels, and each pixel is $15\mu m \times 15\mu m$ in size.
        To make the most use of the screen of SLM, and considering that the beam impinging on the screen is focused to a size of about 4 pixels, we specify that the minimum change unit of $\omega(d)$ is 0.02. The 
        relationship between the spatial spacing $d$ of two path modes and the energy level spacing $\omega$ is set as increasing $\omega$ by 0.02 for every 8-pixel distance increasing of $d$. 
        \begin{figure}[H]
            \centering
            \includegraphics[scale = 0.58]{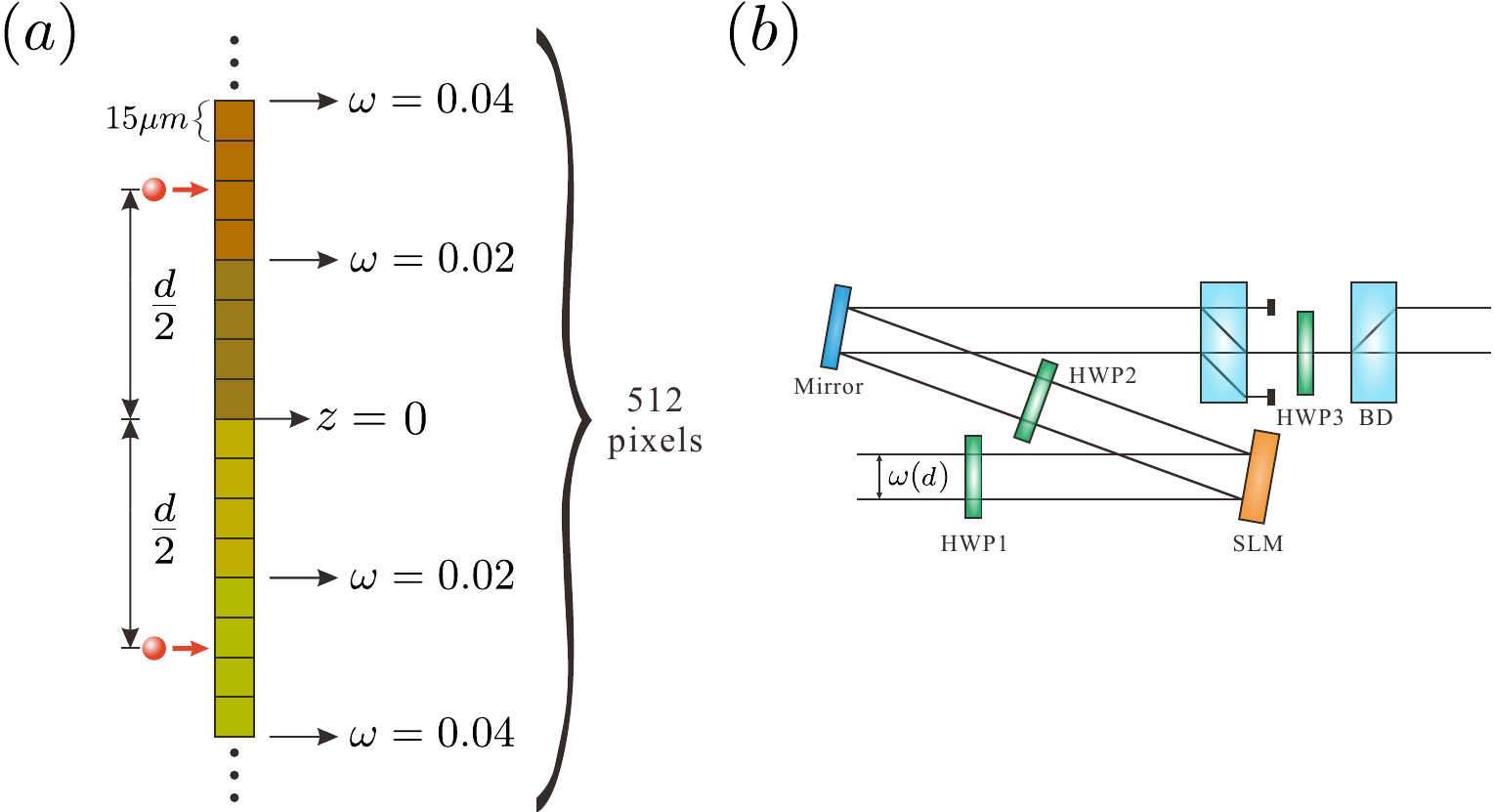}
            \caption{(a) Part of the hologram displayed on SLM. The lower half $(z \le 0)$ is just the image of the upper half plus the image imparted the phase of $\pi$. 
            Every four pixels load the image with the same color. Photonic qubits incident into the same color area and its symmetric areas to $z = 0$ are considered to simulate the working substance with the same energy level spacing. 
            The maximum energy level spacing in this setup is $\omega_{max} = 1.28$. (b) The experimental setup of a thermalizing channel. }
            \label{fig_S1} 
        \end{figure}

        \subsection{2. Thermalizing channel}
        There we introduce the details of the implementation of the thermalizing channel. When a qubit system is in contact with a heat bath and reaches equilibrium, its final state becomes a classical mixed state satisfying the Boltzmann distribution, 
        regardless of the initial state. This process is well described by the thermalizing channel, and its four Kraus operators $\{K_1, K_2, K_3, K_4\}$ can be written as: 
        \begin{equation}
            \begin{aligned}
            K_1 &= \sqrt{p}\begin{pmatrix}1 & 0 \\ 0 & 0 \end{pmatrix},&K_2 = \sqrt{p}\begin{pmatrix}0 & 1 \\ 0 & 0 \end{pmatrix}, \quad K_3 &= \sqrt{1-p}\begin{pmatrix}0 & 0 \\ 0 & 1 \end{pmatrix}, &K_4 = \sqrt{1-p}\begin{pmatrix}0 & 0 \\ 1 & 0 \end{pmatrix}
            \end{aligned}
            \label{ther_cha}
        \end{equation}
        where $p$ is the equilibrium population of the ground state, related to the bath temperature $\beta$ and the qubit energy level spacing $\omega$. The setup is depicted in Fig.~\ref{fig_S1} (b), which consists of an SLM, three HWPs, 
        and two BDs. The specification of BD corresponds to the energy level spacing of the qubit. The hologram $\phi_\beta(z)$ corresponding to the temperature $\beta$ of thermal bath will be loaded on the SLM. 
        Photonic qubit incident on the SLM at the position of $\pm d/2$. Then two different settings are performed to complete the thermalizing channel: 
        in the first setting, angles of HWP1-3 are set to be $22.5^o$, $22.5^o$, $0^o$ respectively, the transformation of this configuration can be expressed as $|V\rangle_{d/2} \rightarrow \cos(\phi_\beta(d/2)/2)|V\rangle_{d/2}$, 
        $|H\rangle_{\textnormal{-}d/2}\rightarrow \cos(\phi_\beta(\textnormal{-}d/2)/2)|H\rangle_{\textnormal{-}d/2}$; in the second setting, angles of HWP1-3 are set to be $22.5^o$, $\textnormal{-}22.5^o$, $45^o$ respectively, and the transformation of this configuration can be expressed as 
        $|V\rangle_{d/2} \rightarrow \sin(\phi_\beta(d/2)/2)|H\rangle_{\textnormal{-}d/2}$, $|H\rangle_{\textnormal{-}d/2}\rightarrow -\sin(\phi_\beta(\textnormal{-}d/2)/2)|V\rangle_{d/2}$. In addition, the upper half of the hologram is applied $\pi$ more than 
        the lower half, i.e. $\phi_\beta(\textnormal{-}z) = \phi_\beta(z) - \pi$, and the coherence between those two encoded states is removed due to the difference between their optical paths. Corresponding them to four Kraus operators, we get 
        \begin{equation}
            \begin{aligned}
            &K_1 : |H\rangle_{\textnormal{-}d/2}\rightarrow \sin(\phi_\beta(d/2)/2)|H\rangle_{\textnormal{-}d/2} \quad &&K_3 : |V\rangle_{d/2} \rightarrow \cos(\phi_\beta(d/2)/2)|V\rangle_{d/2} \notag \\ 
            &K_2 : |V\rangle_{d/2} \rightarrow \sin(\phi_\beta(d/2)/2)|H\rangle_{\textnormal{-}d/2} \quad \quad &&K_4 : |H\rangle_{\textnormal{-}d/2}\rightarrow \cos(\phi_\beta(d/2)/2)|V\rangle_{d/2}
            \end{aligned}
        \end{equation}
        Compared to Eq.~\ref{ther_cha}, we find $p = \sin^2(\phi_\beta(d/2)/2)$. Remember that the qubit system is in equilibrium so that $p = e^{\beta\omega/2}/(e^{\beta\omega/2}+e^{-\beta\omega/2})$. With the functional relationship between $\omega(d)$ and $d$ in the Fig.~\ref{fig_S1} (a) and taking $d$ through the entire SLM screen, 
        the $\phi_\beta(z)$ can be solved. During our experiment, we solved the holograms corresponding to $\beta_1 = 1/2.5$ and $\beta_2 = 1$. The process tomography results of the interaction between different thermal baths and different working substances are shown in Fig.~\ref{fig_S2}. 
        
        \begin{figure}[H]
            \centering
            \includegraphics[scale = 1.05]{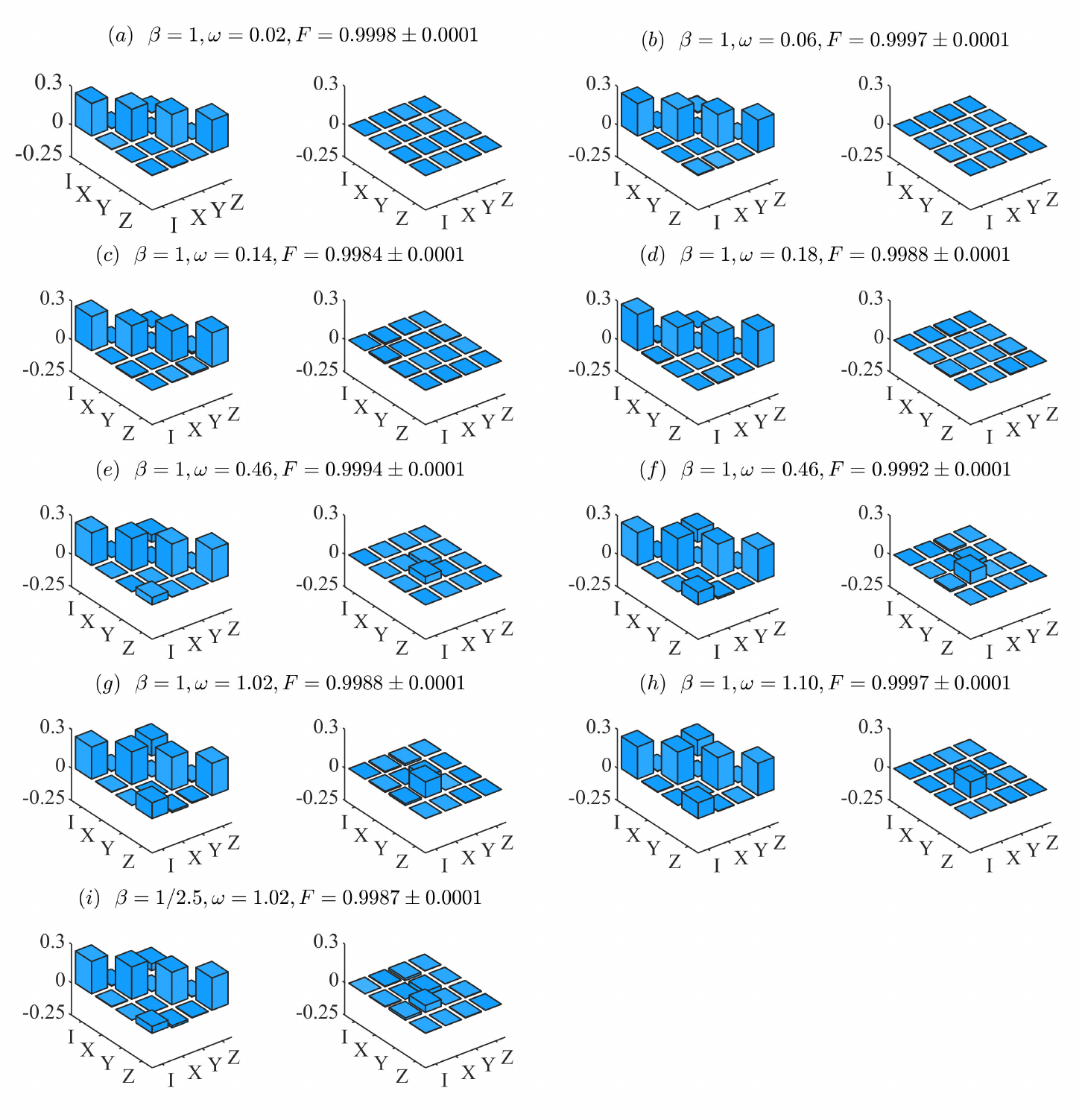}
            \caption{Reconstructed matrices of the results of process tomography, left is the real part and right is the imaginary part in the Pauli basis. The corresponding fidelity is $F$.}
            \label{fig_S2}
        \end{figure}
        
        \subsection{3. Two-qubit polarization measurement}
        The two-qubit polarization measurement is based on the Hong-Ou-Mandel (HOM) effect. Two incident photons overlap with each other at BS. In the case of two photons exiting from different ports of BS, 
        the input state is thus projected onto singlet state $|\psi^-\rangle = (|HV\rangle-|VH\rangle)/\sqrt{2}$. An arbitrary projector $|\varphi_p\rangle\langle\varphi_p|$, in which $|\varphi_p\rangle$ can be 
        written as $|\varphi_p\rangle = U_1U_2(a |HV\rangle - b |VH\rangle)$ according to Schmidt decomposition theorem, is performed by local unitaries and coefficients bias on the singlet projector. 
        The bias of two real coefficients $a$ and $b$ is realized by introducing polarization-dependent loss to one of two photons, which is a Mach-Zehnder (M-Z) interferometer with an HWP placed on each arm. The angle of HWP in the V-arm is $45^o$ while in H-arm 
        is $\theta$. $\theta$ can vary in the range of $0^o$ to $45^o$ (see Fig.2 in the main text). The M-Z interferometer also flips the polarization of the input state, but this can be rectified by the Pauli X operation, which is absorbed into the local unitary 
        performed by three wave plates. In total, the transmission of $V$ polarized state is unit but $H$ is $\sin 2\theta$. The transformation of the bias act on $|\psi^-\rangle$ can be expressed as follows
        \begin{equation}
            |\psi^-\rangle \quad \rightarrow \quad \eta(a |HV\rangle - b |VH\rangle), \quad  \eta = \frac{\sin^2 2\theta + 1}{2}, \quad a = \frac{\sin 2\theta}{\sqrt{1 + \sin^2 2\theta}}, \quad b = \frac{1}{\sqrt{1 + \sin^2 2\theta}}
            \label{mea}
        \end{equation}

        where $\eta$ is the projection efficiency. Together with local unitaries that are performed by three cascade wave plates, we can get $\sqrt{\eta}|\varphi_p\rangle$. In the measurement setup, the unitaries and bias modules are symmetrically placed to BS. 
        In this way, the setup is equivalent to $\eta|\varphi_p\rangle\langle\varphi_p|$. For a measurement setting containing four projectors, $\theta$ is calculated according to Eq.~\ref{mea}, and four projection results are renormalized by their efficiency respectively. 
        By mixing those results we complete the measurement process finally. 

        \begin{figure}[H]
            \centering
            \includegraphics[scale = 0.22]{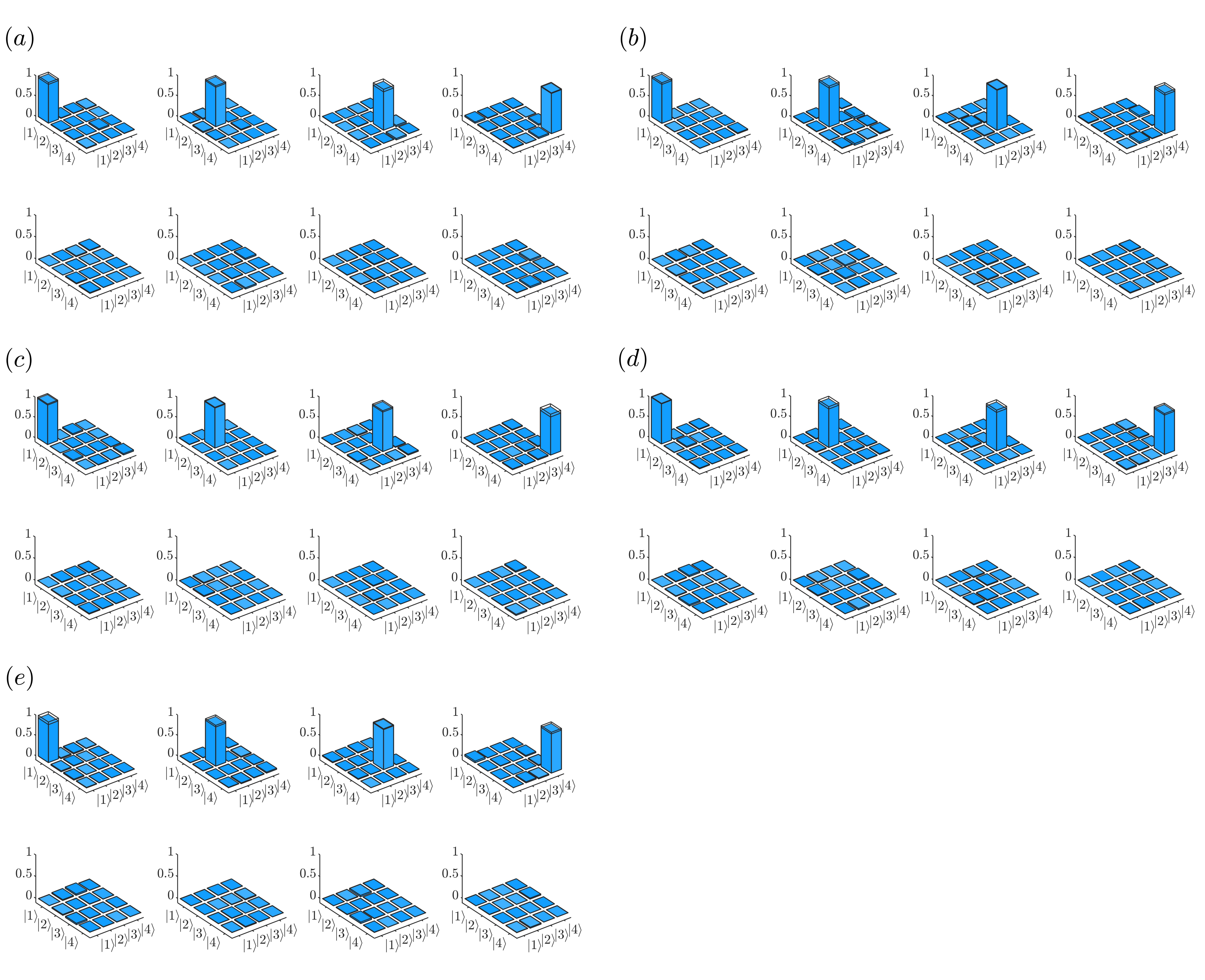}
            \caption{The five results of measurement tomography. In each measurement setting, four top pictures show the real part of four reconstructed projectors and four bottom pictures are the corresponding imaginary part respectively. All the reconstructed matrices are shown 
            in their measurement basis respectively.}
            \label{fig_S3}
        \end{figure}

        To characterize our measurement, we performed measurement tomography for five randomly selected measurement settings, the experimental fidelity is $0.9803 \pm 0.0016, 0.9800 \pm 0.0013, 0.9816 \pm 0.0012, 0.9796 \pm 0.0013, 0.9791 \pm 0.0012$ respectively, the results are shown in Fig.~\ref{fig_S3}. 

        Below we consider the case where the measurement is inefficient. The projective measurement with the basis of Eq.(3) in the main text thus turns into positive operator-valued measurement (POVM). The POVM elements $\{M_k\}_{k=1}^4$ satisfy $M_i^\dagger M_i = \nu|\psi_i\rangle\langle\psi_i|+(1-\nu)I/4$, 
        and $\sum_{k=1}^4 M_k^\dagger M_k = I$. Therefore, we get 
        \begin{equation}
            \begin{aligned}
                |\psi_1 \rangle\langle\psi_1| &\rightarrow M_1 = \frac{\sqrt{1+3\nu}-\sqrt{1-\nu}}{2}|\psi_1\rangle\langle\psi_1| + \frac{\sqrt{1-\nu}}{2} I,  \quad &|\psi_2 \rangle\langle\psi_2| &\rightarrow M_2 = \frac{\sqrt{1+3\nu}-\sqrt{1-\nu}}{2}|\psi_2\rangle\langle\psi_2| + \frac{\sqrt{1-\nu}}{2} I,\\
                |\psi_3 \rangle\langle\psi_3| &\rightarrow M_3 = \frac{\sqrt{1+3\nu}-\sqrt{1-\nu}}{2}|\psi_3\rangle\langle\psi_3| + \frac{\sqrt{1-\nu}}{2} I,  \quad &|\psi_4 \rangle\langle\psi_4| &\rightarrow M_4 = \frac{\sqrt{1+3\nu}-\sqrt{1-\nu}}{2}|\psi_4\rangle\langle\psi_4| + \frac{\sqrt{1-\nu}}{2} I
            \end{aligned}
            \label{noise_basis}
        \end{equation}
        The post-measurement state of the working substance becomes 
        \begin{equation}
            \rho_{\nu}^\prime = \sum_{k=1}^4M_k\rho M_k^\dagger = (\frac{\sqrt{1+3\nu}-\sqrt{1-\nu}}{2})^2\rho^\prime + \frac{(\sqrt{1+3\nu}+\sqrt{1-\nu})\sqrt{1-\nu}}{2}\rho
            \label{rho_noise}
        \end{equation}
        where $\rho$ is the input state and $\rho^\prime$ is the projective post-measurement state. So that only the original projective measurement can change the energy of working substances, the white noise has no effect on the system's energy. 
        For working substances with $\omega_1 = 1.02$ and $\omega_2 = 0.18$, thermal baths at temperatures of $\beta_1 = 1/2.5$ and $\beta_2 = 1$, we plot the energy change of working substances before and after measurement in Fig.~\ref{fig_S4}(a). 
        
        Then we consider the measurement noise of our simulator. This is due to the imperfection of HOM interference at BS. In this scenario, we think that the $\nu$ ratio pairs of photons interfere without defects, where $\nu$ is the interference visibility. But the remaining photons are just divided into two different paths by BS, 
        and only half of them are detected due to post-selection configuration. As a result, the four project operations of measurement become four noisy channels. With the measurement basis of Eq.(3), for working substances with $\omega_1 = 1.02$ and $\omega_2 = 0.18$ and thermal baths at temperatures of $\beta_1 = 1/2.5$ and $\beta_2 = 1$, 
        we decrease $\nu$ from the best value of our simulator. In Fig.~\ref{fig_S4}(b), we see that qubit 2 is transferring less heat from the cold bath. When $\nu$ decreases to the critical visibility $\nu_c \approx 0.44$, the demon can not realize QMC longer. In Fig.~\ref{fig_S4}(c), we plot the critical visibility $\nu_c$, as $\omega_2/\omega_1$ goes down, 
        $\nu_c$ gets smaller and tends to 0 when $\omega_2/\omega_1$ tends to 0. The analysis shows that our simulator is noise-robust, especially when the energy level spacing difference between two qubits is significant.

        \begin{figure}[H]
            \centering
            \includegraphics[scale = 0.63]{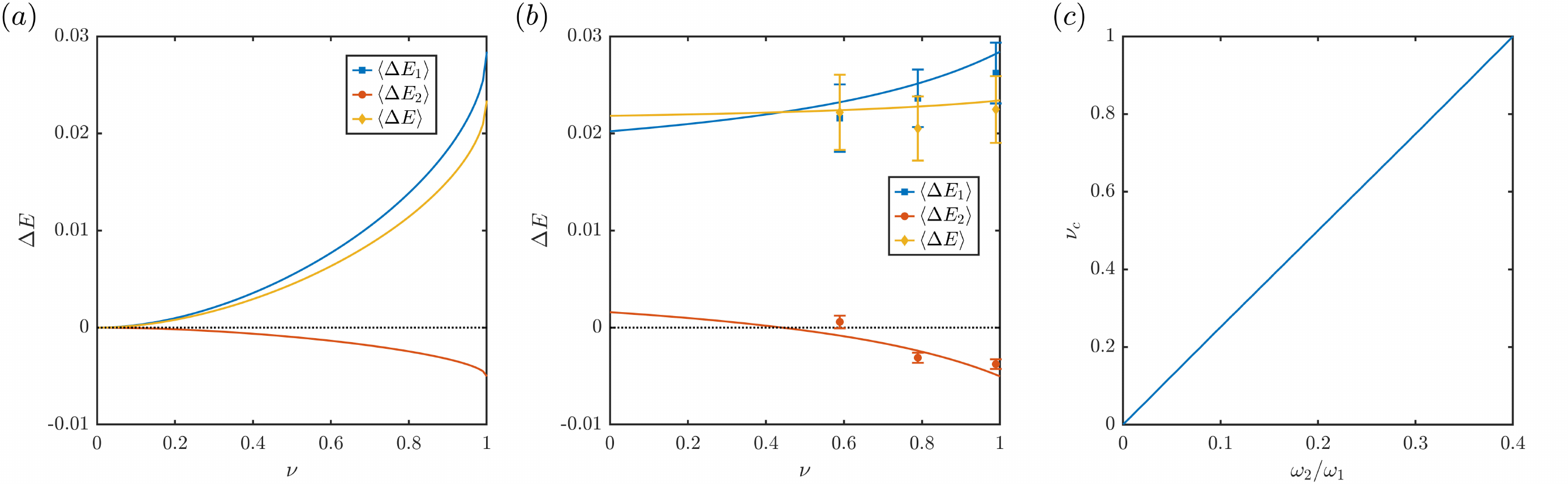}
            \caption{(a) Energy changes of working substances under the white noise of measurement.
            (b) Energy change of working substances under the measurement noise of our simulator. (c) The change in critical visibility $\nu_c$ with respect to $\omega_2/\omega_1$.}
            \label{fig_S4}
        \end{figure}        

        \subsection{4. Other results of measurement cooling engine}
        Suppose the demon has no knowledge about what measurements were performed during the engine strokes. In that case, $[H]$ operation is realized on average, i.e. $\overline{\langle\Delta E_1\rangle} \ge 0$, $\overline{\langle\Delta E_2\rangle} \ge 0$, and $\overline{\langle\Delta E\rangle} \ge 0$. 
        During all the measurement settings of different working substances, our experimental results confirm this conclusion (see Table.~\ref{Ebar}). We also evaluate the frequency of the other three thermodynamic operations on 200 measurement basis, 
        see Fig.~\ref{fig_S5}.
        
        \begin{table}[H]
            \centering
            \setlength{\tabcolsep}{4mm}
            \setstretch{1.5} 
            \begin{tabular}{|c|c|c|c|}
                \hline
                \diagbox{$\overline{\langle E \rangle}$}{$\omega_2$}  & 0.02 & 0.06 & 0.14 \\
                \hline
                $\overline{\langle \Delta E_1 \rangle}$ & 0.09277 $\pm$ 0.00023 & 0.08217 $\pm$ 0.00025 & 0.07593 $\pm$ 0.00028 \\
                \hline
                $\overline{\langle \Delta E_2 \rangle}$ & 0.00009 $\pm$ 0.00001 & 0.00073 $\pm$ 0.00002 & 0.00413 $\pm$ 0.00004 \\
                \hline
                $\overline{\langle \Delta E \rangle}$ & 0.09286 $\pm$ 0.00023 & 0.08290 $\pm$ 0.00024 & 0.08006 $\pm$ 0.00029 \\
                \hline
            \end{tabular}
        \end{table}

        \begin{table}[H]
            \centering
            \setlength{\tabcolsep}{4mm}
            \setstretch{1.5} 
            \begin{tabular}{|c|c|c|c|c|}
                \hline
                \diagbox{$\overline{\langle E \rangle}$}{$\omega_2$}  & 0.18 & 0.46 & 0.86 & 1.10 \\
                \hline
                $\overline{\langle \Delta E_1 \rangle}$ & 0.09159 $\pm$ 0.00027 & 0.08999 $\pm$ 0.00013 & 0.07371 $\pm$ 0.00031 & 0.05855 $\pm$ 0.00028 \\
                \hline
                $\overline{\langle \Delta E_2 \rangle}$ & 0.00853 $\pm$ 0.00005 & 0.05016 $\pm$ 0.00027 & 0.12741 $\pm$ 0.00026 & 0.20197 $\pm$ 0.00027 \\
                \hline
                $\overline{\langle \Delta E \rangle}$ & 0.10012 $\pm$ 0.00027 & 0.14015 $\pm$ 0.00027 & 0.20112 $\pm$ 0.00043 & 0.26052 $\pm$ 0.00039 \\
                \hline
            \end{tabular}
            \caption{Experimental results of $\overline{\langle \Delta E_1 \rangle}$, $\overline{\langle \Delta E_2 \rangle}$, and $\overline{\langle \Delta E \rangle}$. 
            The temperatures of thermal baths are $\beta_1 = 1/2.5$ and $\beta_2 = 1$ respectively, and the energy level spacing of qubit 1 is $\omega_1 = 1.02$.}
            \label{Ebar}
        \end{table}
        
        \begin{figure}[H]
            \centering
            \includegraphics[scale = 0.64]{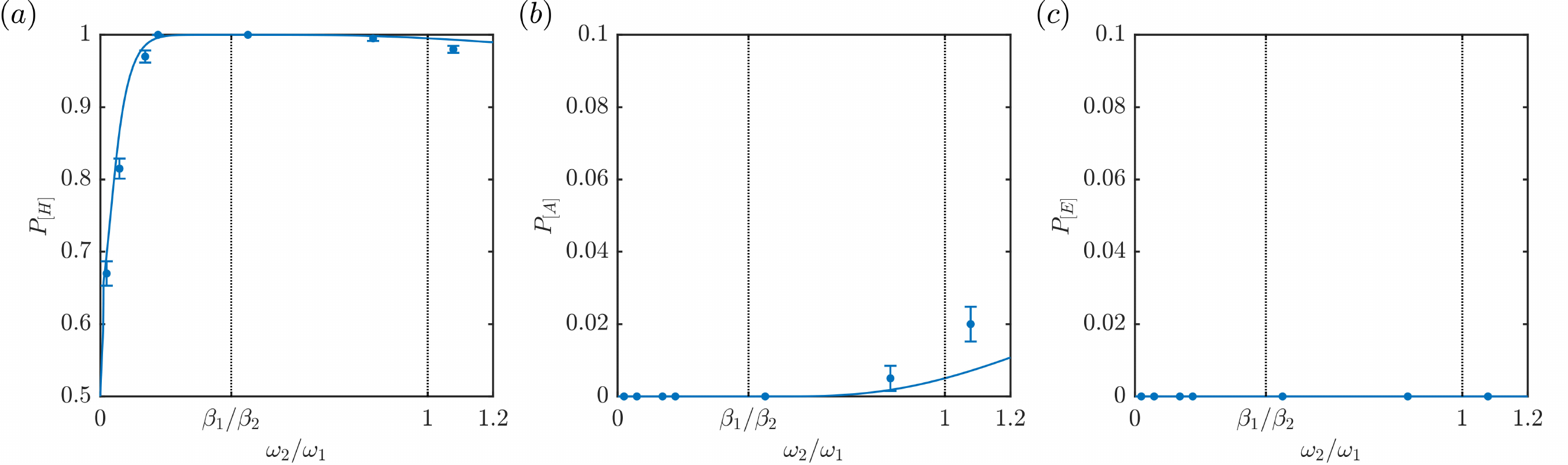}
            \caption{Measured frequency of (a) $[H]$, (b) $[A]$, and (c) $[E]$ operations, when the temperatures of two baths are $\beta_1 = 1/2.5$ and $\beta_2 = 1$ respectively. The probability of realizing $[E]$ operation is extremely small, but $[H]$ operation occurs most frequently.}
            \label{fig_S5}
        \end{figure}

    \clearpage
    \end{widetext}

\end{document}